\documentclass[preprint,12pt]{elsarticle}

\def\beq{\begin{equation}}
\def\eeq{\end{equation}}
\def\beqar{\begin{eqnarray}}
\def\eeqar{\end{eqnarray}}
\def\tb{\tan\beta}

\usepackage{ulem,color}

\usepackage{amssymb}

\newcounter{bla}

\journal{Computer Physics Communications}

\begin{document}

\begin{frontmatter}



\title{ {\bf HDECAY: Twenty{\small ++} Years After} }


\author[a]{Abdelhak Djouadi}
\author[b,e]{Jan Kalinowski$^*$\corref{Jan Kalinowski}}
\author[c]{Margarete M\"{u}hlleitner}
\author[d]{and Michael Spira}

\cortext[author] {Corresponding author. \textit{E-mail address: Jan.Kalinowski@fuw.edu.pl}}
\address[a]{Laboratoire de Physique Th\'eorique, CNRS -- UMR 8627,  
Universit\'e Paris-Sud XI, F-91405 Orsay Cedex, France}
\address[b]{Faculty of Physics, University of Warsaw, ul.\ Pasteura 5, PL-02-093 Warsaw, Poland} 
\address[e]{CERN, Theoretical Physics Department,  CH-1211 Geneva 23, Switzerland }
\address[c]{Institute for Theoretical Physics, Karlsruhe Institute of Technology, 76128 Karlsruhe, Germany}
\address[d]{Paul Scherrer Institut, CH-5232 Villigen PSI, Switzerland}
\begin{abstract}

The program {\tt HDECAY} determines the partial decay widths and
branching ratios of the Higgs bosons within the Standard Model with
three and four  generations of fermions, including the case when the
Higgs couplings are rescaled, a general two--Higgs doublet model where
the Higgs sector is extended and  incorporates five physical states and
its most studied incarnation, the minimal supersymmetric Standard Model
(MSSM). The program addresses all decay channels  including the dominant
higher-order effects such as radiative corrections and multi-body
channels. Since the first launch of the program, more than twenty years
ago, important aspects and new ingredients have been incorporated. In
this update of the program description, some of the developments are
summarized while others are discussed in some detail.

\end{abstract}


\end{frontmatter}



\begin{flushleft} 
Preprint CERN-TH-2017-262 \\
\hspace*{1.5cm} LPT-Orsay-18-04 \\
\hspace*{1.5cm} KA-TP-03-2018 \\
\hspace*{1.5cm} PSI-PR-18-02
\end{flushleft}

\newpage
{\bf NEW VERSION PROGRAM SUMMARY}

\begin{small}

\noindent
{\em Program Title: }   HDECAY                                       \\
{\em Programming language:}      FORTRAN                              \\
{\em Journal reference of previous version: Comp. Phys. Comm. 108 (1998) 56-74.}                  \\
{\em Does the new version supersede the previous version?: YES}   \\
{\em Reasons for the new version: major updates and extensions}\\
   \\

\end{small}



\section{Introduction}
\label{intro}

The FORTRAN code {\tt HDECAY} \cite{Djouadi:1997yw}, released more than
twenty years ago (on arXiv in April 1997)\footnote{Actually, we are not
exactly "twenty years after" \cite{dumas} the first release of the 
program but closer to 42/2 years  and thus, maybe half-way in our
search of the answer to everything  in the universe
\cite{hitchhiker}.}, addresses a crucial issue in the phenomenology of the
Higgs particles which, in the context of the Standard Model (SM) of
particle physics, have been predicted long ago \cite{Higgs} and were
only discovered in 2012 at the CERN Large Hadron Collider (LHC)
\cite{discovery}. Indeed, the strategies for Higgs bosons searches at
high--energy colliders,  such as the LHC,  exploit various Higgs decay
channels. The detection strategies depend not only on the experimental
setup (for instance hadron versus lepton colliders) but also on the
theoretical scenarios: the SM or some of its extensions,  such as the
two-Higgs doublet model (2HDM), or  the Minimal Supersymmetric Standard
Model (MSSM), or variants such as  including a fourth  generation of
fermions, or Higgs particles with rescaled couplings to fermions and
gauge bosons. 

In the SM, the electroweak symmetry is hidden by one doublet scalar
field leading to the existence of one single neutral Higgs boson,
denoted as  $H$ \cite{Higgs}. The Higgs boson couplings to the
3-generation fermions and to gauge bosons are  related to the  masses of
these particles and are thus determined by the symmetry breaking
mechanism. In contrast, the mass $M_H$ of the Higgs boson itself is
undetermined by the model and is known to have the value of $M_H \simeq
125$ GeV only after it was observed by the ATLAS and CMS collaborations
\cite{discovery}. Since this mass value is known, all Higgs couplings,
including the self-couplings, are fixed  and the properties of the $H$
boson, production  cross sections and partial decay widths, are uniquely
determined. 

The situation is more complicated in the  beyond the SM context and, for
instance, additional neutral and  charged Higgs bosons are predicted in 2HDMs 
\cite{HHG,2HDM}, as realized in the MSSM \cite{HHG, 2HD, Djouadi:2005gj}. 
In these models, there are altogether five physical Higgs bosons: 2
CP--even Higgs bosons $h$ and $H$, with $M_h \le M_H$, a CP--odd  or
pseudoscalar $A$ and two charged $H^\pm$ bosons. Either the lighter or
the heavier CP--even Higgs boson can be identified with the one already
observed.
The four  masses $M_h, M_A,
M_H$ and $M_{H^\pm}$, as well as the ratio of the two Higgs field  vacuum
expectation values $\tan\beta=v_2/v_1$  and the mixing angle $\alpha$ that
diagonalises the two CP--even Higgs states, are unrelated in a general 2HDM. In
the MSSM, however, supersymmetry imposes strong constraints on the parameters
and,  in fact, only two of them, usually taken to be $\tan \beta$ and $M_A$, are
independent at tree level.

The MSSM at larger $A$ masses approaches the decoupling regime
\cite{decoupling}  in which the lighter CP--even  $h$ state will have almost
SM--like couplings while the four states $H,A$ and $H^\pm$ become heavy,
degenerate in  mass and decouple from the massive gauge bosons. In the 2HDM, to
cope naturally with the fact that the observed Higgs boson is SM--like, one
invokes the  so--called alignment limit \cite{alignment} in which only one Higgs
doublet gives masses to the $V=W/Z$ bosons. In this case,  the mixing angle
$\alpha$ is such that the Higgs couplings of one of  the CP--even Higgs bosons
are the same as the ones of the SM Higgs state. In this case too, the  second
CP--even state will  no longer couple to massive gauge bosons as also does the
pseudoscalar $A$ in general.  

It was, and still is, of vital importance to have reliable predictions for the
branching ratios of the Higgs boson decays for these theoretical models. The
program {\tt HDECAY} calculates the Higgs boson partial decay widths and the
decay branching ratios within the SM, 2HDM and MSSM scenarios\footnote{The
program has been adapted to very special cases and some versions exist to deal
with specific situations like the MSSM with and without boundary conditions at
the high-energy scale, the {\tt SUSY-HIT} program
\cite{Djouadi:2006bz}, the extension to the
  Next-to-Minimal Supersymmetric SM in {\tt NMSSMCALC} \cite{Baglio:2013iia},  
  the implementation of the Higgs effective Lagrangian
with a linear or non-linear realization of  electroweak symmetry breaking,
{\tt eHDECAY} \cite{Contino:2014aaa}, the
  implementation of the singlet extended SM or 2HDM in {\tt sHDECAY}
  \cite{Costa:2015llh} and {\tt N2HDECAY} \cite{Muhlleitner:2016mzt},
  respectively, and the version for the 
  CP-violating 2HDM {\tt C2HDM\_HDECAY} \cite{Fontes:2017zfn}.}. In
its first version, the main 
features of the program were as follows: 

\begin{itemize}

\item[--] Included are all decay channels that are kinematically allowed and which have
branching ratios larger than $10^{-4}$, i.e.~the loop mediated, the
three-body decay modes and in the MSSM the cascade and the supersymmetric decay 
channels \cite{Djouadi:2005gj,Spira:1997dg,Djouadi:2005gi}.

\item[--] All relevant higher-order QCD corrections to the decays into quark
pairs and to the loop mediated decays into gluons are incorporated
\cite{Spira:1997dg,Djouadi:1995gt}. 

\item[--] Double off--shell decays of the CP--even Higgs bosons into massive
gauge bosons which then decay into four massless fermions, and all
important below-threshold three- and four-body decays are included
\cite{2OFF, Djouadi:1995gv}.

\item[--] In the MSSM, the radiative corrections in the effective
potential approach with full mixing in the stop/sbottom sectors are
incorporated using the renormalization group improved values of the
Higgs masses and couplings thus including the relevant
next--to--leading--order (NLO) and next-to-NLO (NNLO) corrections
\cite{Carena:1995wu,Haber:1996fp,Carena:2000dp}.

\item[--] In the MSSM, all the decays into supersymmetric (SUSY) particles
(neutralinos, charginos, sleptons and squarks including mixing in the stop,
sbottom and stau sectors) when kinematically allowed are calculated 
\cite{Djouadi:1992pu}. The SUSY particles are also
included in the loop mediated $\gamma \gamma, Z\gamma$ and $gg$ decay channels. 

\end{itemize}

The program, written in FORTRAN77, provided a very flexible and
convenient use, fitting to all options of phenomenological relevance.
The basic input parameters, fermion and gauge boson masses and their
total widths, coupling constants and, in the MSSM, soft SUSY-breaking
parameters can be chosen from an input file {\tt hdecay.in}. In this
file several flags allow switching on/off or changing some options [{\it
e.g.} choosing a particular Higgs boson, including/excluding the
multi-body or SUSY decays, or including/excluding some specific
higher-order QCD corrections]. 

All the relevant information is  given in the original publication
\cite{Djouadi:1997yw} to which we refer for details.  However, since the
first release
of the original version of the program some bugs have been fixed, a number of
improvements and new theoretical calculations have been implemented. Earlier
important  modifications were documented, besides those of
Refs.~\cite{Djouadi:2006bz,Contino:2014aaa} where the special extensions {\tt
SUSY-HIT} and  {\tt eHDECAY} were discussed,  in three reports of the Les
Houches Workshops in 1999, 2009 and 2013,  Refs.~\cite{Djouadi:2000gu,
Butterworth:2010ym,Brooijmans:2014eja}. The logbook of all modifications and the
most recent version 6.52 of the program can be found on the web page {\tt
http://tiger.web.psi.ch/hdecay/}. The most important updates are summarized in
the next section. 

\section{The major updates and extensions of the program}


\subsection{Pre-Higgs discovery updates}

Before Higgs boson discovery, most of the modifications of the original
program have been made  in the context of the MSSM. Until 1999 the
following changes have been performed \cite{Djouadi:2000gu}:

\begin{itemize}

\item[--] Added links to the {\tt FeynHiggsFast} routine which provides
the masses and couplings of  the MSSM Higgs bosons up to two--loop order
in the diagrammatic approach \cite{Heinemeyer:1998yj}, and, in the
framework of the {\tt SUSY-HIT} program \cite{Djouadi:2006bz}, to the
{\tt SUSPECT} routine for the renormalisation group evolution and for
the proper electroweak symmetry breaking in the minimal supergravity
model \cite{Djouadi:2002ze}.

\item[--] Implemented Higgs boson decays to a gravitino and neutralino or
char\-gi\-no in gauge mediated SUSY breaking models
\cite{Djouadi:1997gw} and SUSY--QCD corrections to
Higgs boson decays to $q \bar q$ pairs (in particular bottom quarks) 
\cite{Dabelstein:1994hb}.

\end{itemize}

In 2003, a major step was made by providing an interface to the SUSY Les
Houches Accord (SLHA) \cite{Skands:2003cj} and
implementing it properly. This required several transformations of the
corresponding renormalization schemes to the ones used by {\tt HDECAY}.
This option can be switched on and off by appropriate flags in the input
file {\tt hdecay.in}. The output file in the SLHA format can also be
used again as input file for other programs (and {\tt HDECAY} itself). 

Before and at the 2009 Les Houches workshop, the following
modifications, again mainly in the MSSM context, were implemented (some
started to be made in the early 2000). 

\begin{itemize}

\item[--] Improvements of the SUSY--QCD corrections in neutral MSSM
Higgs decays to $b\bar{b}$ \cite{Dabelstein:1994hb} and the resummation
of the $\Delta_b$ effects \cite{Carena:1999py,Guasch:2003cv} up to NNLO
\cite{Noth:2008tw}. The corresponding $\Delta_b$ terms have   also been
included in charged Higgs decays $H^\pm\to tb$.

\item[--] Inclusion of the renormalization-group  improved two-loop
contributions to the  MSSM Higgs self-interactions extending the results
of Ref.~\cite{Carena:1995wu} for the stop and sbottom contributions to
arbitrary mixing parameters and mass splitting; see also
Ref.~\cite{Htriple}.

\item[--] Inclusion of the full mass dependence of the NLO QCD
corrections to the quark and squark loop contributions to photonic Higgs
decays \cite{Djouadi:1990aj, Spira:1995rr}. This was also done in the SM
Higgs case. (The decay widths can  also be used to determine the
production cross sections of Higgs bosons at photon colliders at NLO
QCD.)

\item[--] In the context of the SM, inclusion of electroweak corrections
to the SM Higgs boson decays $H\to W^{(*)}W^{(*)}/Z^{(*)}Z^{(*)} \to 4
f$ in approximate form (including the $WW$ and $ZZ$ thresholds) which
reproduces the full results of Refs.~\cite{Bredenstein:2006rh}
within 1\%. In this context double off-shell effects
have also been extended to the Higgs-mass region above the $WW,ZZ$
thresholds.

\item[--] The full electroweak corrections to the gluonic SM Higgs
decays have been implemented in terms of a grid that is used for
interpolation \cite{Actis:2008ug}. This grid extends up to a Higgs mass
of 1 TeV.

\end{itemize}

After 2009 a few further developments in the context of the MSSM have
been implemented in {\tt HDECAY} before the discovery of the Higgs
particle:

\begin{itemize}

\item[--] The sizeable SUSY--QCD corrections to MSSM Higgs boson decays
into squark pairs \cite{Bartl:1996vh} have been included
supplemented by the improvements concerning the resummation of large
contributions \cite{Accomando:2011jy}. Within the same modification
process the treatment of all squark masses and mixings has been moved
to NLO using the approach and the scheme of Ref.~\cite{Accomando:2011jy}.

\item[--] The MSSM strange Yukawa couplings have been extended to the
inclusion of potentially large $\Delta_s$ effects and their resummation
at the 1-loop level analogous to the $\Delta_b$ effects for the bottom
Yukawa couplings \cite{Carena:1999py, Guasch:2003cv}.

\item[--] Inclusion of running bottom mass and $\Delta_b$ effects in the
top-quark decays $t\to H^\pm b$. For the MSSM (and later for the 2HDM) a
new output file {\tt br.top} is generated that provides the total top
width and the corresponding branching ratios for $t \to Wb, H^\pm b$.

\end{itemize}

\subsection{Post-Higgs discovery modifications: summary}

In the subsequent years and during LHC operation, a large amount of work was
devoted to improve the program. Many of the changes were made in order to meet
the experimental needs and the recommendations of the  LHC Higgs cross section
working group.  By 2013, i.e. a year after the Higgs discovery, very important 
modifications and additions to the program were made. They are summarized below. 

\begin{itemize}

\item[--] Inclusion of the leading SUSY--QCD and electroweak corrections
to all effective down-type fermion Yukawa couplings, i.e.~for the
$\mu,\tau,s$ in addition to the bottom quark according to
Refs.~\cite{Spira:1997dg, Hempfling:1993kv}. In the MSSM the sneutrino
mass parameters of the first two generations are allowed to be different
from the third generation.

\item[--] Inclusion of the two-loop QCD corrections to top decays
\cite{Czarnecki:1998qc}.

\item[--] Inclusion of the full CKM mixing effects in charged Higgs and
top decays.  This required the appropriate extension of the {\tt
hdecay.in} input file.

\item[--] Inclusion of running mass effects and $\Delta_{b/s}$
corrections to the Yukawa couplings in charged Higgs decays into $b$ and
$s$ quarks, where $\Delta_{b/s}$ denotes the leading SUSY-QCD (and
SUSY-electroweak in case of $\Delta_b$) corrections to the effective
bottom/strange Yukawa couplings. Very recently subleading $A_b$ terms
have been included at NNLO in $\Delta_b$ accompanied by their proper
resummation \cite{Guasch:2003cv, Ghezzi:2017enb}. The NNLO results have
been extended to the $\Delta_s$ terms of the MSSM strange Yukawa
couplings \cite{Ghezzi:2017enb}.

\item[--] Addition of the charged Higgs decays $H^+\to t\bar d/ t\bar s/
c\bar d$ including off-shell top quark contributions.

\item[--] Inclusion of charm loop contributions in the Higgs decays
$\phi\to gg$ for the SM and MSSM.

\item[--] Inclusion of the full electroweak corrections to SM Higgs
decays $H\to f\bar f$ \cite{hqqelw} thus removing the approximation used
before.

\item[--] Inclusion of the full NLO mass dependence of SM Higgs decays
into gluons in terms of grids that extend to a Higgs mass of 1 TeV
\cite{Spira:1995rr}.

\item[--] Inclusion of a flag that allows to switch off all electroweak
corrections to SM Higgs decays. This is relevant for consistently using
the best possible predictions of the branching ratios for studies beyond
the SM.

\item[--] The scheme and scale choices of the quark-mass input
parameters have been changed to be in line with the conventions of the
LHCHXWG \cite{deFlorian:2016spz}. This required in particular that the
input values of the file {\tt hdecay.in} have moved to the
$\overline{\rm MS}$ masses $\overline{m}_b(\overline{m}_b)$ for the
bottom and $\overline{m}_c(3~{\rm GeV})$ for the charm quark. The
corresponding bottom pole mass is determined internally by iterating the
N$^3$LO matching relation \cite{msbarpole}
\begin{equation}
{\overline{m}}_{b}(m_{b}^{OS})= \frac{m_{b}^{OS}}{1+\frac{4}{3}
\frac{\alpha_{s}(m_b^{OS})}{\pi} + K_b^{(1)}
\left(\frac{\alpha_s(m_b^{OS})}{\pi}\right)^2 + K_b^{(2)}
\left(\frac{\alpha_s(m_b^{OS})}{\pi}\right)^3}
\end{equation}
at the scale of the bottom pole mass\footnote{Note that this leads to a
slightly different value of the bottom pole mass compared to the
matching at the scale $\overline{m}_b(\overline{m}_b)$ that is, however,
within the corresponding uncertainty band \cite{deFlorian:2016spz}.}
$m_b^{OS}$ where $K_b^{(1)}\sim 12.3$ and $K_b^{(2)}\sim 130.9$. The
charm pole mass is determined from the (renormalon-free) relation
\cite{Bauer:2004ve}
\begin{equation}
m_c^{OS} = m_b^{OS} - 3.41~{\rm GeV} \pm 0.01~{\rm GeV}
\end{equation}
In addition the scale of the input $\overline{\rm MS}$ mass of the
strange quark has been moved to 2 GeV to avoid sizeable non-perturbative
effects when using 1 GeV as the input scale as in former versions of
{\tt HDECAY}. Finally the input values of the $W$ and $Z$ masses and
widths should be chosen as the real parts of the complex poles that are
related to the previous definitions of the physical pole masses
$m_V^{OS}$ and decay widths $\Gamma_V^{OS}$ by
\begin{equation}
m_V - i\Gamma_V = \frac{m_V^{OS} - i\Gamma_V^{OS}}{\sqrt{\displaystyle
1+\left(\frac{\Gamma_V^{OS}}{m_V^{OS}}\right)^2}}
\end{equation}

\item[--] The inclusion of the important option in which the Higgs
couplings to fermions and massive gauge bosons are rescaled by constant
factors in a simplified  effective Lagrangian approach. This also allows
to address the possibilities of fermiophobic or fermiophilic Higgs
states. 

\item[--] The possibility that a fourth generation of SM--like quarks
and leptons is present has been included \cite{Denner:2011vt}. A
significant impact emerges on the loop induced Higgs decays such as
decays into gluons and photons but major changes also occur  in
tree-level decays in which the radiative corrections due  the new
fermions are extremely important. 

\item[--] Extension of {\tt HDECAY} to the general two-Higgs Doublet
model (2HDM) \cite{Harlander:2013qxa}. This required the extension of
the {\tt hdecay.in} input file and the inclusion of several new decay
modes that are not possible within the MSSM. The input file allows to
work with two different sets of input variables for the 2HDM. 

\item[--] Finally, the $h$MSSM option in the supersymmetric case has been
implemented. In this case, the mass of the lightest CP--even MSSM Higgs
boson  $h$, $M_h=125$ GeV, fixes the dominant radiative corrections that
enter the MSSM Higgs boson masses and couplings, leading to a Higgs
sector that can be described, to a good approximation, by only two free
parameters as it was the case at tree-level. 

\end{itemize}

The last four major upgrades are discussed in separate subsections below. 

\subsection{Rescaled Higgs couplings}

In 2013, the program {\tt HDECAY} has been substantially modified (version 6.40)  in order to
cope with the possibility of modified Higgs couplings to fermions and massive
gauge  bosons. This was required by the LHC collaborations which started to
measure precisely the Higgs production cross sections and the decay  branching
ratios, allowing to derive strong constraints on these couplings. To compare the
experimental measurements with the theory predictions in the SM, it was
convenient  to allow for the variation of the different Higgs couplings to the 
other particles in a systematic way.

The inclusion of rescaled Higgs couplings to SM particles has been done 
 according to the simplified effective interaction Lagrangian
\begin{eqnarray}
{\cal L}_{int} & \ni & -\left\{\sum_\psi c_\psi m_\psi \bar\psi \psi 
+ 2 c_W m_W^2 W^{+\mu} W^-_\mu  + c_Z m_Z^2 Z^{\mu} Z_\mu\right\}
\frac{H}{v} \nonumber \\
& + & \left\{ \frac{\alpha_s}{8\pi} c_{gg} G^{a\mu\nu}G^a_{\mu\nu}
+ \frac{\alpha}{8\pi} c_{\gamma\gamma} F^{\mu\nu}F_{\mu\nu}
+ \frac{\sqrt{\alpha \alpha_2}}{4\pi} c_{Z\gamma}
F^{\mu\nu}Z_{\mu\nu}\right\} \frac{H}{v}
\end{eqnarray}
where $G^{a\mu\nu}$, $F^{\mu\nu}$ and $Z^{\mu\nu}$ are the field strength
tensors of the gluon, photon and $Z$-boson fields. The couplings $\alpha_s$,
$\alpha$ and $\alpha_2$ are the strong, electromagnetic (in the Thompson limit) and 
SU(2) isospin ($g^2 = 4 \pi \alpha_2$) couplings, respectively, $v$ is the
Higgs vacuum expectation value and $H$ the Higgs boson field. The 
novel point-like couplings of the Higgs boson to gluons, photons and $Z$ bosons
affect the Higgs decays $H\to gg/\gamma\gamma/Z\gamma$. Electroweak
corrections are only kept in the SM part of the individual decay amplitudes,
i.e.~the parts for $c_\psi=c_W=c_Z=1$ and $c_{gg}=c_{\gamma\gamma}=c_{Z\gamma}
=0$, while QCD corrections have been included in all parts of the decays widths,
since the dominant parts factorize. This approach deviates from the general
addition of dimension-six operators as pursued in Ref.~\cite{Contino:2013kra}
where additional tensor structures have been added at the dimension-six level.

The above rescaling of the Higgs couplings modifies e.g.~the Higgs decay
widths into quarks as
\begin{equation}
\Gamma(H\to q\bar q) = \frac{3G_FM_H}{4\sqrt{2}\pi}
\overline{m}_q^2(M_H) c_b \left\{ c_b + \delta_{elw} \right\} \left\{
1+\delta_{QCD} + \frac{c_t}{c_b} \delta_t \right\}
\end{equation}
where $\delta_{elw}$ denotes the electroweak corrections \cite{hqqelw},
$\delta_{QCD}$ the pure QCD corrections
\cite{nearthreshold, abovethreshold, chetyrkin},
$\delta_t$ the top-quark induced QCD correction
\cite{abovethreshold6}, and $\overline{m}_q$ is  the running
$\overline{\rm MS}$ quark mass  at the scale of the Higgs mass.

The gluonic Higgs decay, with  the novel tensor
structure involving the point-like coupling factor $c_{gg}$, is given by
\begin{eqnarray}
\Gamma(H\to gg)  &=& \frac{G_F\alpha_s^2 M_H^3}{36\sqrt{2}\pi^3}
\Bigg[ 
 \bigg| \sum_{Q=t,b,c} c_Q \, A_Q\left(\tau_Q\right)
 \bigg|^2  c_{eff}^2 \, \kappa_{soft}  \nonumber \\
&& + \delta_{elw} \left( \sum_{Q,Q'=t,b,c} c_Q \, A_Q\left(\tau_Q\right)
A_Q^*\left(\tau_{Q'}\right) \right)
 c_{eff}^2 \, \kappa_{soft}  \nonumber \\
&& + 2\, \mathrm{Re}\!\left( \sum_{Q=t,b,c} c_Q \,
A^*_Q\left(\tau_Q\right)
\frac{3}{2} c_{gg} \right) c_{eff} \, \kappa_{soft}  
    + \left|\frac{3}{2} c_{gg}\right|^2
    \kappa_{soft}    \nonumber  \\
&& + \sum_{Q,Q'=t,b} c_Q\, A^*_Q\left(\tau_Q\right)
c_{Q'}\, A_Q\left(\tau_{Q'}\right) \kappa^{NLO} (\tau_Q,\tau_{Q'})
\Bigg] \, ,
\end{eqnarray}
where $\tau_Q = 4 m_Q^2 /M_H^2$ and $\delta_{elw}$ denotes the
electroweak corrections \cite{Actis:2008ug, Aglietti:2004nj}.
The loop function $A_Q(\tau_Q)$ is
normalized to unity for large quark masses and can be found in
Ref.~\cite{Spira:1995rr}. The contributions $c_{eff}$ and $\kappa_{soft}$ denote
the QCD corrections originating from the effective Lagrangian in the heavy top
quark limit,
\begin{equation}
{\cal L}_{eff} = c_{eff}~\frac{\alpha_s}{12\pi} G^{a\mu\nu}G^a_{\mu\nu}
\frac{H}{v}
\end{equation}
and the residual corrections due to diagrams involving gluon exchange and
light-quark contributions, respectively. They are included up to the
next-to-next-to-next-to-leading order (NNNLO)
\cite{Inami:1982xt, Chetyrkin:1997iv}.
At  NLO, they are given by
\cite{Inami:1982xt},
\begin{equation}
c_{eff} = 1 + \frac{11}{4}\, \frac{\alpha_s}{\pi}\; , \qquad \qquad
\kappa_{soft} = 1 + \left( \frac{73}{4} - \frac{7}{6} N_F \right)
\frac{\alpha_s}{\pi}
\end{equation}
with $N_F=5$ light quark flavours.  Finally $\kappa^{NLO}$ represents the finite
top and bottom mass effects at NLO beyond the limit of heavy quarks, i.e.~beyond
the terms contained in $c_{eff}$ and $\kappa_{soft}$ \cite{Spira:1995rr}.

All other Higgs decay modes are treated analogously in the case of
rescaled Higgs couplings.


\subsection{The fourth generation fermion option}

In the four-generation fermion Standard Model (SM4), available since
version 4.0 of the {\tt HDECAY} code, additional corrections to
tree--level Higgs decays into fermions and massive gauge bosons arise
from 4th-generation fermion loops. In the case of the $H\to WW,ZZ$
decays, these corrections  appear in the $HWW/HZZ$ vertices, the $W/Z$
self-energies, and the renormalization constants. Since the
4th-generation fermions are expected to be very heavy, their  Yukawa
couplings are large and they dominate the total corrections. Numerically
the NLO corrections  amount to about $-50\%$ to $-90\%$ in many cases
and depend only weakly on the Higgs mass \cite{Denner:2011vt}. In the
large fermion mass limit, the leading contribution can be absorbed into
effective $HWW/HZZ$ interactions via the Lagrangian
\beq
{\cal L}_{HVV} = \sqrt{\sqrt{2} G_F}~H \left[ 2 M_W^2 W^{+\mu} W^-_\mu (1 + \delta_{W}^{\mathrm{tot}}) + M_Z^2 Z_\mu Z^\mu (1 + \delta_{Z}^{\mathrm{tot}}) \right],
\eeq
where $W,Z,H$ denote the fields for the respective states. The higher-order corrections are contained in the factors $\delta_V^{\mathrm{tot}}$ which,
up to two-loop order, read
\beq
\delta_V^{\mathrm{tot}(1)} = \delta_u^{(1)} + \delta_V^{(1)}, \qquad
\delta_V^{\mathrm{tot}(2)} = \delta_u^{(2)} + \delta_V^{(2)} + \delta_u^{(1)} \delta_V^{(1)}.
\eeq
The one-loop expressions for a single $\mathrm{SU}(2)$ doublet of heavy
fermions with masses $m_A$, $m_B$ read \cite{Chanowitz:1978uj,Djouadi:1997rj}
\beq
\delta_u^{(1)} = N_c X_A \left[ \frac{7}{6} (1 + x) + \frac{x}{1-x} \ln x\right], \qquad \delta_V^{(1)} = - 2 N_c X_A (1+x),
\label{eq:delta1}
\eeq
where $x = m_B^2 / m_A^2$,  $X_A = G_F m_A^2/(8 \sqrt{2} \pi^2)$, and $N_c=3$ or
$1$ for quarks or leptons, respectively. The results for the two-loop
corrections $\delta_V^{\mathrm{tot}(2)}$ can be found in \cite{Kniehl:1995ra}
for the QCD corrections of ${\cal O}(\alpha_s G_F m_f^2)$ and in
\cite{Djouadi:1997rj} for the EW ones of ${\cal O}(G_F^2 m_f^4)$.  The
corrected partial decay width is then given by
\beq
\Gamma_{\mathrm{NLO}} \approx \Gamma_{\mathrm{LO}} \left[1 + 
\delta_\Gamma^{(1)} + \delta_\Gamma^{(2)}\right] =
\Gamma_{\mathrm{LO}} \left[1 + 2 \delta_V^{\mathrm{tot}(1)} +
(\delta_V^{\mathrm{tot}(1)})^2 + 2 \delta_V^{\mathrm{tot}(2)}\right].
\eeq

In the case of Higgs decays into SM fermions, the decay widths $\Gamma(
H\to f\bar{f})$ in   the {\tt  HDECAY} code  include, besides the SM
corrections, the 4th generation approximate NLO and NNLO EW corrections
in the heavy-fermion limit according to \cite{Djouadi:1997rj} and mixed
NNLO EW/QCD corrections  according to \cite{Kniehl:1995ra}. They
originate from the wave-function renormalization of the Higgs boson and
are thus universal for all fermion species. The leading one-loop part is
given by $\delta_u^{(1)}$ above. Numerically the EW one-loop correction
to the partial widths amounts to a few tens of  percent,  
while the two-loop EW and QCD correction contributes an additional few
percent.  
These corrections are assumed to factorize since the approximate
expressions emerge as corrections to the effective Lagrangian after
integrating out the heavy fermion species.  Thus, {\tt HDECAY}
multiplies the relative SM4 corrections with the full corrected SM3
(usual SM with three generations) result including QCD and EW
corrections. The scale of the strong coupling $\alpha_s$ is set to the
average mass of the heavy 4th generation quarks according to the
appropriate matching scale of the effective Lagrangian.
 
Turning to the loop induced decay modes $H\to gg, \gamma\gamma,
Z\gamma$, {\tt HDECAY} includes the NNNLO QCD corrections of the SM in
the limit of a heavy top quark \cite{Spira:1995rr,Inami:1982xt,
Chetyrkin:1997iv}, applied to the results
including the heavy-quark loops. For $H\to gg$, while at NNLO the exact
QCD corrections in SM4 \cite{Anastasiou:2010bt} are
included in this limit, at NNNLO the relative SM3 corrections are added
to the relative NNLO corrections and multiplied by the LO result
including the additional quark loops.  In addition the full NLO EW
corrections \cite{Passarino:2011kv} have been included in factorized
form, since the dominant part of the QCD corrections emerges from the
gluonic contributions on top of the corrections to the effective
Lagrangian in the limit of heavy quarks. 

{\tt HDECAY} includes the full NLO QCD corrections to the decay mode
$H\to \gamma\gamma$ supplemented by the additional contributions of the
4th-generation quarks and charged leptons according to
\cite{Djouadi:1990aj, Spira:1995rr}.  Extending the techniques used for
$H \to gg$ in \cite{Passarino:2011kv}, the exact amplitude for $H \to
\gamma\gamma$ was included up to NLO (two-loop level) as in
Ref.~\cite{Denner:2011vt}. It required particular attention as in many
scenarios the correction is negative and larger than unity. This is due
the fact that at LO the cancellation between  the $W$ and the fermion
loops is stronger in SM4 than in SM3 so that the LO  result is
suppressed more.  Furthermore, the NLO corrections are strongly enhanced
for ultra-heavy fermions. In such a case, a proper estimate must also
include the next term in the expansion \cite{Passarino:2011kv} which is
included in the grid implemented in {\tt HDECAY} for $H\to\gamma\gamma$.

Finally, the decay mode $H\to \gamma Z$ is treated at LO only, since the
NLO QCD corrections within the SM3 are known to be small
\cite{Spira:1991tj} and can thus safely be neglected. The EW corrections
in SM3 as well as in SM4 are unknown which implies a sizeable
theoretical uncertainty within SM4, since large cancellations between
the $W$ and fermion loops emerge at LO.


\subsection{The two-Higgs doublet model extension}

{\tt HDECAY} version 6.0 has been extended to include the computation of the
Higgs boson decay widths in the framework of the 2HDM \cite{Harlander:2013qxa}.
The most general (CP-conserving) version of the 2HDM with a softly-broken
$Z_2$-symmetry, i.e.~ type I--IV models, has been implemented. For the
input parameters, to be specified in the input file {\tt hdecay.in},
the user can choose between two different options given by 
\begin{eqnarray}
\begin{array}{ll}
\mbox{the ratio of the vacuum expectation values:} & \tan\beta \\
\mbox{the mass parameter squared:} & M_{12}^2 \; (\mathrm{GeV}^2) \\
\mbox{the quartic couplings of the Higgs potential:} & \lambda_1, ... , 
\lambda_5 \end{array}
\nonumber          
\end{eqnarray} 
or by a more physical basis
\begin{eqnarray}
\begin{array}{ll}
\mbox{the ratio of the vacuum expectation values:} & \tan\beta \\
\mbox{the mass parameter squared:} & M_{12}^2 \; (\mathrm{GeV}^2) \\
\mbox{the CP--even Higgs mixing angle:} & \alpha \\
\mbox{the mass values of the five Higgs bosons:} & M_h, M_H, M_A,
M_{H^\pm}  \; (\mathrm{GeV})\\
\end{array}
\nonumber          
\end{eqnarray}  
Furthermore, one can choose between the four 2HDM types by setting the
corresponding flag accordingly.  From these input parameters {\tt
HDECAY} calculates the couplings which are needed in the computation of
the decay widths. With the appropriate coupling replacements according
to the various 2HDMs, the decay widths are the same as for the MSSM
Higgs boson decays, which are already included in the program. Only the
SUSY particle contributions in the loop-mediated decays and the decays
into SUSY particles as well as higher order corrections due to SUSY
particle loops have been turned off for the 2HDM case.  

The decay widths of the 2HDM Higgs bosons are usually calculated at LO
in the 2HDM parameters. Unlike the case of the SM with a light Higgs or
the MSSM, there is no automatic protection in the 2HDM against
arbitrarily large quartic couplings, which may lead to a violation of
perturbativity of the couplings and tree-level unitarity. This should be
kept in mind when calculating decay widths involving Higgs
self--couplings.  The program also tests for vacuum stability,
perturbative unitarity and the compatibility with the $S$ and $T$
parameters and gives out a warning if these are not fulfilled.

Higher-order SM EW corrections do not factorize from the LO result and
cannot be readily included for the 2HDM. The higher-order EW corrections
to all relevant 2HDM Higgs boson decays have only become available
recently \cite{Krause:2016oke}, and not been implemented in {\tt HDECAY}
yet.  This introduces unavoidable uncertainties, which can be estimated
from the size of the known EW SM corrections to be up to 5--10\% for
several partial decay widths. Differences of this magnitude compared to
the most precise values for the SM are therefore expected even in the
decoupling/alignment limit. Note, however, that the corrections can be
much larger for Higgs-to-Higgs decays, where they can be parametrically
enhanced.

A consistent comparison of 2HDM predictions in the decoupling/alignment
limit to SM values furthermore requires that the limit is taken properly
so that no residual 2HDM effects are present, e.g.~from $H^\pm$ loop
contributions to $h \!\to \! \gamma\gamma/Z\gamma$. Using the physical
Higgs mass basis as an example, SM-like decays for the lightest 2HDM $h$
boson can be achieved by choosing $M_h\sim 125$~GeV,
$\sin(\beta\!-\!\alpha)\!=\!1$, $M_{H,A,H^\pm} \! \gg \! v$, and
$M_{12}^2$ such that $g_{hH^+H^-}=0$.

Unlike EW corrections, many of the QCD corrections (which typically are
numerically significant) do factorize, and can therefore be taken over
directly from the corresponding SM or MSSM calculations. The widths for
SM-type Higgs boson decays to quark and vector bosons pairs are obtained at LO from
their SM equivalents by rescaling the vertices with the corresponding
2HDM factors. The loop-mediated decay to gluons also proceeds as in the
SM, with the appropriate rescaling of the Yukawa couplings. For the
remaining decays it is necessary in addition to take 2HDM-specific
contributions (which cannot be assumed to be small) into account.

In {\tt HDECAY}, the implemented decay widths and higher order corrections are
specified in the following. 

\begin{itemize}

\item[--] Decays into quark pairs: The QCD corrections factorize and can
be taken over from the SM case. For the neutral Higgs decays the fully
massive NLO corrections near threshold \cite{nearthreshold}  and the
massless ${\cal O} (\alpha_s^4)$ corrections far above threshold
\cite{abovethreshold, chetyrkin, abovethreshold6} are included in {\tt
HDECAY}. They are calculated in terms of running quark masses and strong
coupling to resum large logarithms. The QCD corrections to $H^\pm$
decays have been taken from \cite{Hpm}.  The EW corrections cannot be
adapted  from the SM case and are ignored. For the decays of the heavier
neutral Higgs bosons into a top quark pair, in {\tt HDECAY} off-shell
decays below threshold have been implemented as well as for the decays
of a charged Higgs boson into a top-bottom quark pair
\cite{Djouadi:1995gv}. 

\item[--] Decays into gluons: The QCD corrections to the neutral Higgs
boson decays into gluons, a loop-induced process already at leading
order, can be taken over from the SM, respectively, the MSSM.  They have
been included up to N$^3$LO in {\tt HDECAY} in the limit of heavy top
quarks.  While for the SM at NLO the full quark mass dependence
\cite{Spira:1995rr} is available, for the 2HDM the corrections have been
taken into account in the limit of heavy-quark loop particle masses
\cite{Spira:1995rr, Inami:1982xt, Chetyrkin:1997iv, nloggqcd}.

\item[--] Decays into $\gamma\gamma, Z\gamma$: The decay to a photon
pair is loop-mediated, with the two most important SM contributions
being due to the top quark and $W$ boson loops. In the 2HDM, there is
also a $H^+$ contribution which becomes numerically significant in some
cases. The bottom loop becomes relevant in scenarios with enhanced
bottom Yukawa couplings.  In the pseudoscalar case only heavy charged
fermion loops contribute.  The neutral Higgs boson decays into a photon
pair have been implemented at NLO QCD including the full mass dependence
for the quarks \cite{Djouadi:1990aj, Spira:1995rr, hgagalim}.  The loop
induced decays of scalar Higgs bosons into $Z\gamma$ are mediated by
$W$, charged Higgs and heavy charged fermion loops, while the
pseudoscalar decays proceed only through charged fermion loops. The QCD
corrections to the quark loops are numerically small \cite{Spira:1991tj}
and have not been taken into account in {\tt HDECAY}.

\item[--] Decays into massive gauge bosons: The decay widths of the
scalar Higgs bosons into massive gauge bosons $\phi \to V^{(*)}V^{(*)}$
are the same as the SM decay width after replacing the SM Higgs coupling
to gauge bosons with the corresponding 2HDM Higgs coupling. The option
of double off-shell decays \cite{2OFF} has been included in {\tt
HDECAY}. The pseudoscalar Higgs boson does not decay into massive gauge
bosons at tree level. 

\item[--] Decays into Higgs boson pairs: The heavier Higgs particles can
decay into a pair of lighter Higgs bosons. This is a feature of the 2HDM
which does not exist in the SM. Due to more freedom in the mass
hierarchies compared to the MSSM case, the following Higgs-to-Higgs
decays are possible and have been taken into account in {\tt HDECAY},
\begin{eqnarray}
h \to AA^{(*)} \,, \qquad H \to hh^{(*)} \,, \qquad H \to AA^{(*)}  \;.
\end{eqnarray}
Moreover the decays into a charged Higgs boson pair are
possible\footnote{Note that in type II and IV (flipped), certain decays
are already kinematically closed due to the lower bound of $M_{H^\pm} >
580$ GeV on the charged Higgs boson mass \cite{Misiak:2017bgg}.},
\begin{eqnarray}
h \to H^+ H^- \,, \quad \quad H \to H^+ H^- \;.
\end{eqnarray}
All decays are calculated at leading order using the tree-level
expressions of the 2HDM trilinear couplings.  The contributions from
final states with an off-shell scalar or pseudoscalar, which can be
significant, have been included in {\tt HDECAY} \cite{Djouadi:1995gv}.
It is important to note that these partial widths can be very large for
parameter points that do not respect the requirements of perturbativity
and tree-level unitarity.

-- Decays into gauge and Higgs bosons: The Higgs boson decays into a
gauge and a Higgs boson \cite{Djouadi:2005gj, Spira:1997dg},
which have been implemented in {\tt HDECAY}  including
the possibility of off-shell gauge bosons \cite{Djouadi:1995gv},
are in particular
\begin{eqnarray}
\begin{array}{lllllllll}
h &\to& A Z^{(*)} \,, & h &\to& H^\pm W^{\mp (*)}\,, \\
H &\to& A Z^{(*)} \,, & H &\to& H^\pm W^{\mp (*)}\,, \\
A &\to& h Z^{(*)} \,, &A &\to& H Z^{(*)} \,, & A &\to&
H^\pm W^{\mp (*)} \,,  \\  
H^\pm &\to& h W^{\pm (*)}\,, & H^\pm &\to& H W^{\pm (*)}\,, & H^\pm
&\to& A W^{\pm (*)} \;.
\end{array}
\end{eqnarray}
They have been implemented at leading order and include the contributions of
off-shell $W$ and $Z$ bosons below threshold \cite{Djouadi:1995gv}.

\end{itemize}

\subsection{The $h$MSSM scenario}

As mentioned earlier, in the MSSM, only two parameters are needed to describe
the Higgs sector at tree-level. These are in general taken to be $M_A$ and
$\tb$. Nevertheless, when the radiative corrections \cite{RC1} are included in
the Higgs sector, in particular the dominant loop contributions from the top and
stop quarks that have strong couplings to the Higgs bosons \cite{RC2}, many
supersymmetric parameters will enter the game. This is for instance the case of
the SUSY scale,  taken to be the geometric  average of the two stop masses $M_S=
\sqrt {m_{\tilde t_1} m_{\tilde t_2} }$,  the stop/sbottom trilinear couplings
$A_{t/b}$ or the higgsino mass $\mu$ (other corrections, that involve the
gaugino mass parameters $M_{1,2,3}$ for instance are rather small). 

In particular, the radiative corrections in the CP--even neutral Higgs
sector are extremely  important and shift the value of the lightest $h$
boson mass from the tree--level value  $M_h \leq M_Z \cos2\beta \leq
M_Z$ to the value $M_h=125$ GeV that has been measured experimentally.
In the current-eigenstate basis of the Higgs fields $\Phi_1, \Phi_2$,
the CP--even Higgs mass matrix including corrections can be written as:
\beq
M_{S}^2=M_{Z}^2 \! \left( \!
\begin{array}{cc}
  c^2_\beta \! & \! -s_\beta c_\beta \\ -s_\beta c_\beta \! & \! s^2_\beta \\
\end{array}  \! \right) \! + \! M_{A}^2 \! \left( \!
\begin{array}{cc}  s^2_\beta \! & \! -s_\beta c_\beta \\
 -s_\beta c_\beta \! & \! c^2_\beta \\
\end{array} \! \right) \!
\! + \!
\left( \!
\begin{array}{cc}
 \Delta {\cal M}_{11}^2 &  \Delta {\cal M}_{12}^2 \\
 \Delta {\cal M}_{12}^2 &\Delta {\cal M}_{22}^2 \\
\end{array}
\! \right)
\label{matrix-hMSSM}
\eeq
where we have used the short--hand notation $c_\beta \equiv \cos\beta ,
s_\beta \equiv \sin\beta$  and  the radiative corrections  are captured
by a general $2\times 2$ matrix $\Delta {\cal M}_{ij}^2$. The neutral
CP--even Higgs boson masses and the mixing angle $\alpha$ that
diagonalises the $h,H$  states can then easily be derived, $H=
\Phi_1^0 \cos\alpha + \Phi_2^0 \sin\alpha$ and $h= -\Phi_1^0 \sin\alpha
+ \Phi_2^0 \cos\alpha$, where $\Phi_{1,2}^0$ denote the neutral CP--even
components of the physical Higgs fields in the current-eigenstate basis.
In the $2\times 2$ matrix $\Delta {\cal M}^2$, only the $\Delta{\cal
M}^{2}_{22}$ entry is in fact relevant in most cases (in particular if
$\mu$ is small). It involves the by far dominant stop--top sector
correction \cite{RC2}, 
\beq 
\Delta{\cal M}^{2}_{22} \approx \Delta M_h^2|^{t/\tilde{t}}_{\rm 1loop} \sim 
\frac{3m_t^4}{2\pi^2v^2} \bigg[ \log \frac{M_{S}^2} {m_{t}^2} +\frac{X_{t}^{2}}{M_{S}^{2}}
- \frac{X_{t}^{4} }{12 M_{S}^{4}} \bigg] \;,
\eeq 
where $M_S$ is the SUSY scale and $X_t=A_t-\mu/\tb$ the stop mixing
parameter.  Hence, one can write $\Delta{\cal M}^{2}_{22} \gg
\Delta{\cal M}^{2}_{11}, \Delta{\cal M}^{2}_{12}$ in general. It has
been advocated~\cite{Djouadi:2013uqa, Bagnaschi:2015hka} that in this
case,  one can simply trade $\Delta {\cal M}^{2}_{22}$ for the known
$M_h$ value using 
\beq
\Delta {\cal M}^{2}_{22}= \frac{M_{h}^2(M_{A}^2  + M_{Z}^2 -M_{h}^2) -
  M_{A}^2 M_{Z}^2 c^{2}_{2\beta} } { M_{Z}^2 c^{2}_{\beta}  +M_{A}^2
  s^{2}_{\beta} -M_{h}^2} \;. 
\label{m22-hMSSM}
\eeq
One can then simply write $M_H$ and $\alpha$ in terms of
$M_A,\tb$ and $M_{h}$:
\begin{eqnarray}
{\mbox{$h$MSSM}}:~~ 
\begin{array}{l} 
M_{H}^2 = \frac{(M_{A}^2+M_{Z}^2-M_{h}^2)(M_{Z}^2
  c^{2}_{\beta}+M_{A}^2
s^{2}_{\beta}) - M_{A}^2 M_{Z}^2 c^{2}_{2\beta} } {M_{Z}^2
c^{2}_{\beta}+M_{A}^2
s^{2}_{\beta} - M_{h}^2} \\
\ \ \  \alpha = -\arctan\left(\frac{ (M_{Z}^2+M_{A}^2) c_{\beta}
    s_{\beta}} {M_{Z}^2 
c^{2}_{\beta}+M_{A}^2 s^{2}_{\beta} - M_{h}^2}\right)
\end{array} \;.
\label{wide} 
\end{eqnarray}
In the case of the $H^\pm$  masses, the radiative corrections are
small at large enough $M_A$ and one has to a good
approximation \cite{Frank:2013hba}
\begin{eqnarray}
M_{H^\pm} \simeq \sqrt { M_A^2 + M_W^2}.
\end{eqnarray}
This is the $h$MSSM approach which has been shown to provide a good
approximation of the MSSM Higgs sector. In this $h$MSSM, the MSSM Higgs
sector can be again described with only the two parameters $\tb$  and
$M_A$ as the loop corrections are fixed by the value of $M_h$. Another
advantage of this approach is that it allows to describe the low $\tb$
region of the MSSM which was overlooked as for SUSY scales of order 1
TeV, values $\tb <3$ were excluded because they lead to an $h$ mass that
is smaller than 125 GeV. The price to pay is that for such  low $\tb$
values, one has to assume $M_S \gg 1$ TeV and, hence, that the model is
fine-tuned. Moreover, care has to be taken not to enter regimes for
small values of $\tan\beta$ that cannot be accommodated with the MSSM as
pointed out in Ref.~\cite{Bagnaschi:2015hka}.

The couplings of the CP--even $h$ and $H$ to fermions and vector bosons
are given in terms of the angle $\alpha$ which, including the radiative
correction, is fixed by the $h$MSSM relations above. Additional direct
corrections as $\Delta_b$ should in principle enter the Higgs couplings
but because $M_S$ is taken to be very large, they are assumed to have a
small impact in the $h$MSSM and are ignored. This, however, strongly
depends on the size of the $\mu$ parameter and should be taken with
caution for large values of $\tan\beta$. Another important set of
couplings are the Higgs self-couplings and in the $h$MSSM, they are
again given in terms of $\beta$ and $\alpha$, with the latter fixed by
$\tb$, $M_A$ and $M_h$ as in Eq.~(\ref{wide}), but contain additional
genuine radiative corrections that can be derived from the input
parameters, too, since they are related to $\Delta{\cal M}^{2}_{22}$.

The calculation of the Higgs branching ratios within the $h$MSSM are
performed by {\tt HDECAY} starting with version  6.40. The program takes
$M_h$ as input and obtains  $M_H$ and $\alpha$ from the $h$MSSM
prescriptions.  For the decays, the $h$MSSM mode of {\tt HDECAY}
implements:  N$^4$LO-QCD corrections to the decays to quark pairs; LO
results for the decays to lepton pairs and for the decays involving
massive gauge bosons, both on-shell and off-shell; a LO calculation of
the decays to Higgs-boson pairs, both on-shell and off-shell, using
effective $h$MSSM couplings such as the $Hhh$ coupling in particular.
The triple Higgs couplings are an important  issue \cite{Htriple} which
needs some further studies in the $h$MSSM and some preliminary results
recently appeared in \cite{Chalons:2017wnz}.



\newpage \section{The input file} In the following we list the input
parameters of the {\tt hdecay.in} input file along with some
explanations\footnote{The choices of the flag {\tt MODEL} refer to the
following References: {\tt MODEL = 1} \cite{Carena:1995wu}, {\tt MODEL = 2}
\cite{Carena:1995bx}, {\tt MODEL = 3} \cite{Haber:1996fp}, {\tt MODEL =
4} \cite{Heinemeyer:1998yj}, {\tt MODEL = 10} \cite{Djouadi:2013uqa}.}.

\begin{small}
\begin{verbatim}

   SLHAIN: =0: READ FROM hdecay.in
           =1: READ SUSY LES HOUCHES ACCORD INPUT (slha.in)

  SLHAOUT: =0: WRITE BR TABLES
           =1: WRITE SUSY LES HOUCHES ACCORD OUTPUT (slha.out)

  COUPVAR: =0: NO VARIATION OF HIGGS COUPLINGS
           =1: VARIATION OF HIGGS COUPLINGS       (ONLY FOR SM)

    HIGGS: =0: CALCULATE BRANCHING RATIOS OF SM HIGGS BOSON
           =1: CALCULATE BRANCHING RATIOS OF MSSM h BOSON
           =2: CALCULATE BRANCHING RATIOS OF MSSM H BOSON
           =3: CALCULATE BRANCHING RATIOS OF MSSM A BOSON
           =4: CALCULATE BRANCHING RATIOS OF MSSM H+ BOSON
           =5: CALCULATE BRANCHING RATIOS OF ALL MSSM HIGGS BOSONS

  OMIT ELW =0: INCLUDE FULL ELECTROWEAK CORRECTIONS (SM)
           =1: OMIT ALL ELECTROWEAK CORRECTIONS (SM)

      SM4: =0: CALCULATE USUAL BRANCHING RATIOS
           =1: HIGGS WITH 4TH GENERATION (SETS HIGGS, FERMPHOB = 0)

 FERMPHOB: =0: CALCULATE USUAL BRANCHING RATIOS
           =1: FERMIOPHOBIC HIGGS (SETS HIGGS = 0)

     2HDM: =0: CALCULATE USUAL BRNCHING RATIOS
           =1: 2HDM (SETS HIGGS = 5)

    MODEL: USE SPECIFIC SUBROUTINE FOR MSSM HIGSS MASSES AND COUPLINGS
           =1: CARENA ET AL., NUCL. PHYS. B461 (1996) 407 (SUBHPOLE)
           =2: CARENA ET AL., PHYS. LETT. B355 (1995) 209 (SUBH)
           =3: HABER ET AL.
           =4: HEINEMEYER ET AL., HEP-PH/0002213 (FEYNHIGGSFAST1.2.2)
           =10: hMSSM

 TGBET:     TAN(BETA) FOR MSSM
 MABEG:     START VALUE OF M_A FOR MSSM AND M_H FOR SM
 MAEND:     END VALUE OF M_A FOR MSSM AND M_H FOR SM
 NMA:       NUMBER OF ITERATIONS FOR M_A
 MHL:       LIGHT SCALAR HIGGS MASS FOR hMSSM (MODEL = 10)
 ALS(MZ):   VALUE FOR ALPHA_S(M_Z)
 MSBAR(2):  MSBAR MASS OF STRANGE QUARK AT SCALE Q=2 GEV
 MCBAR(3):  CHARM MSBAR MASS AT SCALE Q=3 GEV
 MBBAR(MB): BOTTOM MSBAR MASS AT SCALE Q=MBBAR
 MT:        TOP POLE MASS
 MTAU:      TAU MASS
 MMUON:     MUON MASS
 ALPH:      INVERSE QED COUPLING
 GF:        FERMI CONSTANT
 GAMW:      W WIDTH
 GAMZ:      Z WIDTH
 MZ:        Z MASS
 MW:        W MASS
 VTB:       CKM PARAMETER |V_TB|
 VTS:       CKM PARAMETER |V_TS|
 VTD:       CKM PARAMETER |V_TD|
 VCB:       CKM PARAMETER |V_CB|
 VCS:       CKM PARAMETER |V_CS|
 VCD:       CKM PARAMETER |V_CD|
 VUB:       CKM PARAMETER |V_UB|
 VUS:       CKM PARAMETER |V_US|
 VUD:       CKM PARAMETER |V_UD|
 GG_ELW:    SCENARIO OF THE ELW. CORRECTIONS TO H -> GG (4TH GENERATION)
 MTP:       TOP' MASS    (4TH GENERATION)
 MBP:       BOTTOM' MASS (4TH GENERATION)
 MNUP:      NU' MASS     (4TH GENERATION)
 MEP:       E' MASS      (4TH GENERATION)

  2HDM models

 TYPE:      1 (type I), 2 (type II), 3 (lepton-specific), 4 (flipped)
 PARAM:     1 (masses), 2 (lambda_i)
 TGBET2HDM: TAN(BETA)
 ALPHA_H:   MIXING ANGLE IN THE CP-EVEN NEUTRAL HIGGS SECTOR
 MHL:       MASS OF THE LIGHT CP-EVEN HIGGS BOSON
 MHH:       MASS OF THE HEAVY CP-EVEN HIGGS BOSON
 MHA:       MASS OF THE CP-ODD HIGGS BOSON
 MH+-:      MASS OF THE CHARGED HIGGS BOSONS
 LAMBDA1:   2HDM lambda parameter
 LAMBDA2:   2HDM lambda parameter
 LAMBDA3:   2HDM lambda parameter
 LAMBDA4:   2HDM lambda parameter
 LAMBDA5:   2HDM lambda parameter
 M_12^2:    PARAMETER M12 SQUARED
 
 SUSYSCALE: SCALE FOR SUSY BREAKING PARAMETERS
 1ST AND 2ND GENERATION:
 MSL1:      SUSY BREAKING MASS PARAMETERS OF LEFT HANDED SLEPTONS 
 MER1:      SUSY BREAKING MASS PARAMETERS OF RIGHT HANDED SLEPTONS 
 MQL1:      SUSY BREAKING MASS PARAMETERS OF LEFT HANDED SUPS
 MUR1:      SUSY BREAKING MASS PARAMETERS OF RIGHT HANDED SUPS
 MDR1:      SUSY BREAKING MASS PARAMETERS OF RIGHT HANDED SDOWNS 
 3RD GENERATION:
 MSL:       SUSY BREAKING MASS PARAMETERS OF LEFT HANDED STAUS 
 MER:       SUSY BREAKING MASS PARAMETERS OF RIGHT HANDED STAUS 
 MSQ:       SUSY BREAKING MASS PARAMETERS OF LEFT HANDED STOPS
 MUR:       SUSY BREAKING MASS PARAMETERS OF RIGHT HANDED STOPS
 MDR:       SUSY BREAKING MASS PARAMETERS OF RIGHT HANDED SBOTTOMS 
 AL:        STAU TRILINEAR SOFT BREAKING TERMS 
 AU:        STOP TRILINEAR SOFT BREAKING TERMS
 AD:        SBOTTOM TRILINEAR SOFT BREAKING TERMS
 MU:        SUSY HIGGS MASS PARAMETER
 M2:        GAUGINO MASS PARAMETER
 MGLUINO:   GLUINO POLE MASS

 ON-SHELL: =0: INCLUDE OFF_SHELL DECAYS H,A --> T*T*, A --> Z*H,
               H --> W*H+,Z*A, H+ --> W*A, W*H, T*B
           =1: EXCLUDE THE OFF-SHELL DECAYS ABOVE

 ON-SH-WZ: =0: INCLUDE DOUBLE OFF-SHELL PAIR DECAYS PHI --> W*W*,Z*Z*
           =1: INCLUDE DOUBLE OFF-SHELL PAIR DECAYS PHI --> W*W*,Z*Z*
               BELOW THRESHOLD, BUT ON-SHELL PAIR DECAYS ABOVE
           =-1: INCLUDE ONLY SINGLE OFF-SHELL DECAYS PHI --> W*W,Z*Z
                BELOW THRESHOLD, BUT ON-SHELL PAIR DECAYS ABOVE

 IPOLE:    =0 COMPUTES RUNNING HIGGS MASSES (FASTER) 
           =1 COMPUTES POLE HIGGS MASSES 

 OFF-SUSY: =0: INCLUDE DECAYS (AND LOOPS) INTO SUPERSYMMETRIC PARTICLES
           =1: EXCLUDE DECAYS (AND LOOPS) INTO SUPERSYMMETRIC PARTICLES

 INDIDEC:  =0: PRINT OUT SUMS OF CHARGINO/NEUTRALINO/SFERMION DECAYS
           =1: PRINT OUT INDIVIDUAL CHARGINO/NEUTRALINO/SFERMION DECAYS

 NF-GG:    NUMBER OF LIGHT FLAVORS INCLUDED IN THE GLUONIC DECAYS 
            PHI --> GG* --> GQQ (3,4 OR 5)
           
 IGOLD:    =0: EXCLUDE DECAYS INTO GRAVITINO + GAUGINO
           =1: INCLUDE DECAYS INTO GRAVITINO + GAUGINO

 MPLANCK:  PLANCK MASS FOR DECAYS INTO GRAVITINO + GAUGINO
 MGOLD:    GRAVITINO MASS FOR DECAYS INTO GRAVITINO + GAUGINO

          RESCALING OF COUPLINGS

 ELWK:    = 0: Include elw. corrections only for SM part
          = 1: Include elw. corrections in all rescalings of couplings
 CW:      RESCALING FACTOR OF HWW COUPLING
 CZ:      RESCALING FACTOR OF HZZ COUPLING
 Ctau:    RESCALING FACTOR OF HTAUTAU COUPLING
 Cmu:     RESCALING FACTOR OF HMUMU COUPLING
 Ct:      RESCALING FACTOR OF HTT COUPLING
 Cb:      RESCALING FACTOR OF HBB COUPLING
 Cc:      RESCALING FACTOR OF HCC COUPLING
 Cs:      RESCALING FACTOR OF HSS COUPLING
 Cgaga:   POINT-LIKE H-GAMMA-GAMMA COUPLING
 Cgg:     POINT-LIKE HGG COUPLING
 CZga:    POINT-LIKE H-Z-GAMMA COUPLING

   4th generation fermions   

 Ctp:     RESCALING FACTOR OF HT'T' COUPLING
 Cbp:     RESCALING FACTOR OF HB'B' COUPLING
 Cnup:    RESCALING FACTOR OF HNU'NU' COUPLING
 Cep:     RESCALING FACTOR OF HE'E' COUPLING
\end{verbatim}
\end{small}

In the next section we will give sample output files for the various
models implemented in {\tt HDECAY}. The input file that we use to
generate the outputs is given here. We will then later only indicate
the changes of those parameters in the input file that are relevant for the
specific model. The input file {\tt hdecay.in} reads
\begin{small}
\begin{verbatim}
SLHAIN   = 0
SLHAOUT  = 0
COUPVAR  = 0
HIGGS    = 0  
OMIT ELW = 0
SM4      = 0
FERMPHOB = 0
2HDM     = 0
MODEL    = 1
TGBET    = 30.D0
MABEG    = 125.D0
MAEND    = 1000.D0
NMA      = 1  
********************* hMSSM (MODEL = 10) *********************************
MHL      = 125.D0
**************************************************************************
ALS(MZ)  = 0.1180D0
MSBAR(2) = 0.095D0
MCBAR(3) = 0.986D0
MBBAR(MB)= 4.180D0
MT       = 173.2D0
MTAU     = 1.77682D0
MMUON    = 0.1056583715D0
1/ALPHA  = 137.0359997D0
GF       = 1.1663787D-5
GAMW     = 2.08430D0
GAMZ     = 2.49427D0
MZ       = 91.15348D0
MW       = 80.35797D0
VTB      = 0.9991D0
VTS      = 0.0404D0
VTD      = 0.00867D0
VCB      = 0.0412D0
VCS      = 0.97344D0
VCD      = 0.22520D0
VUB      = 0.00351D0
VUS      = 0.22534D0
VUD      = 0.97427D0
********************* 4TH GENERATION *************************************
  SCENARIO FOR ELW. CORRECTIONS TO H -> GG (EVERYTHING IN GEV):
  GG_ELW = 1: MTP = 500    MBP = 450    MNUP = 375    MEP = 450
  GG_ELW = 2: MBP = MNUP = MEP = 600    MTP = MBP+50*(1+LOG(M_H/115)/5)

GG_ELW   = 1
MTP      = 500.D0
MBP      = 450.D0
MNUP     = 375.D0
MEP      = 450.D0
************************** 2 Higgs Doublet Model *************************
  TYPE: 1 (I), 2 (II), 3 (lepton-specific), 4 (flipped)
  PARAM: 1 (masses), 2 (lambda_i)

PARAM    = 1
TYPE     = 2 
********************
TGBET2HDM= 1.29775D0
M_12^2   = 82857.8D0
******************** PARAM=1:
ALPHA_H  = -0.684653D0
MHL      = 125.09D0
MHH      = 453.87D0
MHA      = 591.552D0
MH+-     = 613.93D0
******************** PARAM=2:
LAMBDA1  = 0.989175D0
LAMBDA2  = 0.734211D0
LAMBDA3  = 6.42606D0
LAMBDA4  = -3.83528D0
LAMBDA5  = -2.94533D0
**************************************************************************
SUSYSCALE= 500.D0
MU       = 200.D0
M2       = 200.D0
MGLUINO  = 1500.D0
MSL1     = 1000.D0
MER1     = 1000.D0
MQL1     = 1000.D0
MUR1     = 1000.D0
MDR1     = 1000.D0
MSL      = 1000.D0
MER      = 1000.D0
MSQ      = 1000.D0
MUR      = 1000.D0
MDR      = 1000.D0
AL       = 1607.D0
AU       = 1607.D0
AD       = 1607.D0
ON-SHELL = 0
ON-SH-WZ = 0
IPOLE    = 0
OFF-SUSY = 0
INDIDEC  = 0
NF-GG    = 5
IGOLD    = 0
MPLANCK  = 2.4D18
MGOLD    = 1.D-13
******************* VARIATION OF HIGGS COUPLINGS *************************
ELWK     = 0
CW       = 1.D0
CZ       = 1.D0
Ctau     = 1.D0
Cmu      = 1.D0
Ct       = 1.D0
Cb       = 1.D0
Cc       = 1.D0
Cs       = 1.D0
Cgaga    = 0.D0
Cgg      = 0.D0
CZga     = 0.D0
********************* 4TH GENERATION *************************************
Ctp      = 0.D0
Cbp      = 0.D0
Cnup     = 0.D0
Cep      = 0.D0
\end{verbatim}
\end{small}

\section{The output files}
We will give exemplary output files for the SM, the 2HDM and the MSSM
based on the input file given above with the respective changes for
these models given below. 
\subsection{The Standard Model}
The input file can be taken over without any changes. It 
leads to the following {\tt br.sm1} and {\tt br.sm2} output files given by
\begin{small}
\begin{verbatim}
  MHSM      BB      TAU TAU     MU MU       SS         CC         TT 
______________________________________________________________________
 125.000  0.5811   0.6259E-01 0.2172E-03 0.2239E-03 0.2886E-01  0.000 
\end{verbatim}
\end{small}
and
\begin{small}
\begin{verbatim}
   MHSM      GG      GAM GAM    Z GAM        WW     ZZ        WIDTH
______________________________________________________________________
 125.000 0.8164E-01 0.2265E-02 0.1529E-02 0.2152 0.2634E-01 0.4096E-02
\end{verbatim}
\end{small}
respectively, for the branching ratios of the SM Higgs with mass
$M_{H_{\mathrm{SM}}}=125$~GeV 
into the bottom-quark, tau- and muon-pair, strange-, charm- and top-quark
pair final states as well as into gluon, photon, $Z\gamma$ and massive
gauge boson final states. The last entry in {\tt br.sm2} is the total
width in GeV. 

\subsection{The 2HDM}
For the 2HDM example we chose a scenario compatible with all
relevant theoretical and experimental constraints \cite{Muhlleitner:2017dkd} and also
implies a strong first order phase transition as required by 
baryogenesis \cite{Basler:2016obg}. It induces a mass spectrum
where the lightest CP-even Higgs boson is the SM-like Higgs state.
The input parameters are specified
in the above input file, and only the following two parameters need to
be changed to produce the 2HDM output files:
\begin{small}
\begin{verbatim}
HIGGS    = 5
2HDM     = 1
\end{verbatim}
\end{small}
For the given scenario the input is the 'physical' one via the Higgs masses
and the mixing angles and {\tt PARAM} is set equal to 1. If {\tt PARAM} is
set equal to 2 the $\lambda_i$'s as 
given in the above input file lead to the same results. 
The output for the branching ratios is given in the files {\tt br.xy\_2HDM}
with $x=l,h,a,c$ for the light and heavy CP-even $h$ and
$H$ states, for the CP-odd Higgs $A$ and the charged boson $H^\pm$,
respectively. The index $y$ counts the output files of each Higgs
boson. The three output files {\tt br.l1\_2HDM},  {\tt br.l2\_2HDM} and
{\tt br.l3\_2HDM} for the SM-like Higgs with mass
$m_h\! =\! 125.09$~GeV read\footnote{Note that in
  the present version of {\tt HDECAY} no SLHA output files are 
  provided in the 2HDM case.}:
\begin{small}
\begin{verbatim}
  MHL       BB     TAU TAU     MU MU        SS         CC        TT 
______________________________________________________________________
 125.090  0.6080  0.6542E-01  0.2316E-03  0.2294E-03 0.2653E-01 0.000
\end{verbatim}
\end{small}
\begin{small}
\begin{verbatim}
  MHL        GG        GAM GAM     Z GAM       WW       ZZ 
______________________________________________________________________
 125.090  0.7041E-01  0.2126E-02  0.1458E-02  0.2005  0.2507E-01
\end{verbatim}
\end{small}
\begin{small}
\begin{verbatim}
  MHL        AA      Z A    W+- H-+   H+ H-     WIDTH 
______________________________________________________________________
 125.090   0.000    0.000    0.000    0.000   0.4248E-02
\end{verbatim}
\end{small}
As can be inferred from these files the $h$ boson behaves SM-like.
For $H$ we obtain the three output files {\tt br.h1\_2HDM},  {\tt
br.h2\_2HDM} and {\tt br.h3\_2HDM}:
\begin{small}
\begin{verbatim}
   MHH      BB       TAU TAU     MU MU        SS         CC      TT 
______________________________________________________________________
 453.870 0.1869E-02 0.2595E-03 0.9176E-06 0.6831E-06 0.3675E-04 0.9807
\end{verbatim}
\end{small}
\begin{small}
\begin{verbatim}
   MHH          GG      GAM GAM     Z GAM         WW         ZZ 
______________________________________________________________________
 453.870    0.3781E-02 0.1094E-04 0.2425E-05 0.3372E-02 0.1595E-02
\end{verbatim}
\end{small}
\begin{small}
\begin{verbatim}
   MHH         hh        AA     Z A     W+- H-+    H+ H-      WIDTH 
______________________________________________________________________
 453.870   0.8341E-02  0.000   0.000    0.000      0.000      5.837  
\end{verbatim}
\end{small}
The heavy Higgs boson with mass above 350~GeV dominantly decays into a
top-quark pair. It can also decay into a pair of lighter Higgs
bosons. The branching ratio is rather small, however.
The three output files {\tt br.a1\_2HDM}, {\tt br.a2\_2HDM} and {\tt
  br.a3\_2HDM} of the pseudoscalar Higgs boson are given by:
\begin{small}
\begin{verbatim}
  MHA      BB       TAU TAU     MU MU       SS         CC       TT 
______________________________________________________________________
591.552 0.7844E-03 0.1121E-03 0.3965E-06 0.2950E-06 0.1365E-04 0.8873
\end{verbatim}
\end{small}
\begin{small}
\begin{verbatim}
   MHA        GG       GAM GAM     Z GAM        Z h        Z H
______________________________________________________________________
 591.552  0.2648E-02  0.7803E-05  0.2312E-05  0.2341E-02  0.1068
\end{verbatim}
\end{small}
\begin{small}
\begin{verbatim}
   MHA       W+- H-+     WIDTH 
______________________________________________________________________
 591.552      0.000      18.41   
\end{verbatim}
\end{small}
Also the pseudoscalar dominantly decays into a top-quark
pair. However, the decay into the $Z$ boson and the heavy Higgs
boson $H$ contributes with 10\% and is a prime example of a
beyond-the-SM decay. 
For the charged Higgs boson the three generated output files {\tt br.c1\_2HDM},
{\tt br.c2\_2HDM} and {\tt br.c3\_2HDM} read:
\begin{small}
\begin{verbatim}
   MHC       BC       TAU NU      MU NU       SU         CS       TB 
______________________________________________________________________
 613.930 0.1192E-05 0.1023E-03 0.3619E-06 0.1296E-07 0.1145E-04 0.7794
\end{verbatim}
\end{small}
\begin{small}
\begin{verbatim}
   MHC         CD         BU         TS         TD   
______________________________________________________________________
 613.930   0.6001E-06 0.8503E-08 0.1273E-02 0.5863E-04
\end{verbatim}
\end{small}
\begin{small}
\begin{verbatim}
   MHC           hW       HW         AW      WIDTH 
______________________________________________________________________
 613.930     0.2381E-02 0.2168   0.3495E-06  20.93
\end{verbatim}
\end{small}
The first two files contain the branching ratios for the fermionic final
states, and the last one the charged Higgs branching ratios into
Higgs-gauge boson final states, which can become significant here in
the $HW$ case. In addition the top-quark branching ratios and
total width are given in the file {\tt br.top}
\begin{small}
\begin{verbatim}
   MHC         W+- B      H+- B      WIDTH
______________________________________________________________________
 613.930      1.000      0.000      1.336
\end{verbatim}
\end{small}
For this 2HDM scenario the top quark decays entirely into $Wb$
final states.

\subsection{The MSSM}
In order to generate the output file for the MSSM with a SM-like Higgs
boson mass close to 125~GeV, we have to change in the input file given
above
\begin{small}
\begin{verbatim}
HIGGS    = 5
MABEG    = 1000.D0
\end{verbatim}
\end{small}
The scenario that corresponds to the MSSM is the $m_h^{mod+}$
scenario defined in Ref.~\cite{benchmark}. It induces a mass spectrum
with the lightest CP-even Higgs boson having a mass close to 125~GeV,
namely $m_h = 122.644$~GeV. The branching ratios, given in {\tt br.l1}
and {\tt br.l2}, are
\begin{small}
\begin{verbatim}
  MHL       BB     TAU TAU     MU MU       SS         CC        TT 
______________________________________________________________________
 122.644  0.6419  0.6686E-01 0.2367E-03 0.2348E-03 0.2905E-01  0.000
\end{verbatim}
\end{small}
\begin{small}
\begin{verbatim}
   MHL      GG      GAM GAM      Z GAM      WW      ZZ        WIDTH
______________________________________________________________________
 122.644 0.7593E-01 0.2195E-02 0.1263E-02 0.1628 0.1953E-01 0.3993E-02
\end{verbatim}
\end{small}
The file {\tt br.ls} includes the branching ratios into the SUSY
particle final states, which are all kinematically closed, however, so that we do not
give {\tt br.ls} separately here. Being in the decoupling limit with
the chosen large pseudoscalar mass of 1~TeV, the branching ratios are
close to those of a SM Higgs boson with same mass. For $H$, the
branching ratios into SM particle and Higgs boson final states listed
in {\tt br.h1}, {\tt br.h2}, {\tt br.h3} amount to:
\begin{small}
\begin{verbatim}
  MHH      BB    TAU TAU     MU MU       SS         CC         TT 
______________________________________________________________________
 1000.02 0.4215 0.7915E-01 0.2799E-03 0.1627E-03 0.3208E-07 0.2062E-02
\end{verbatim}
\end{small}
\begin{small}
\begin{verbatim}
   MHH        GG       GAM GAM     Z GAM         WW          ZZ 
______________________________________________________________________
 1000.02  0.5130E-04  0.3036E-06  0.2152E-07  0.9420E-05  0.4652E-05
\end{verbatim}
\end{small}
\begin{small}
\begin{verbatim}
   MHH        hh         AA         Z A      W+- H-+   H+ H-  WIDTH
_____________________________________________________________________
 1000.02  0.4391E-04  0.8720E-23  0.4605E-19  0.000   0.000   23.78
\end{verbatim}
\end{small}
Due to the large value of $\tan\beta=30$ the branching ratio into
$b\bar{b}$ dominates over the one into $t\bar{t}$.  While the decay into
the light Higgs boson pair is kinematically open, it is very small. The
decay into $AA$ is far off-shell and hence tiny\footnote{Tiny negative
value of the branching ratio may arise due to an artefact of the finite
accuracy of the implemented expressions in the Fortan code and should be
ignored, i.e.~identified with zero.}.  The decays into SUSY particles
are given in {\tt br.hs}:
\begin{small}
\begin{verbatim}
TB= 30.0000     M2= 200.000     MU= 200.000     MSQ= 1000.00    
C1=148.714 C2= 266.081 N1= 88.414 N2=152.084 N3= 210.462 N4= 265.541
MST1= 856.771     MST2= 1101.73     MSUL= 975.811     MSUR= 976.706  
MSB1= 973.103     MSB2= 983.796     MSDL= 979.235     MSDR= 977.688  
TAU1= 997.129 TAU2=1004.930 NL= 997.925 EL= 1001.15 ER= 1000.92 
NL1= 997.93

  MHH     CHARGINOS  NEUTRALS   SLEPTONS   SQUARKS  GRAVITINO+GAUGINO
______________________________________________________________________
 1000.02    0.3027     0.1940      0.000      0.000      0.000
\end{verbatim}
\end{small}
The kinematically allowed decays into SUSY particles, on the
other hand, are important with branching ratios into charginos and
neutralinos of about 30\% and 20\%, respectively. Note that the SUSY
particle branching ratios given in the output files sum up all the final states of the
same SUSY particle type. In {\tt br.hs} also
the masses of the SUSY particles are repeated in the output for convenience.
The branching ratios of the pseudoscalar are summarized in the output
files {\tt br.a1}, {\tt br.a2} and {\tt br.as}:
\begin{small}
\begin{verbatim}
  MHA      BB    TAU TAU     MU MU       SS         CC          TT 
______________________________________________________________________
 1000.00 0.4216 0.7916E-01 0.2799E-03 0.1627E-03 0.3078E-07 0.2158E-02
\end{verbatim}
\end{small}
\begin{small}
\begin{verbatim}
   MHA         GG       GAM GAM     Z GAM        Z HL      WIDTH 
______________________________________________________________________
 1000.00   0.9014E-04  0.5085E-06  0.3919E-07  0.9280E-05  23.78
\end{verbatim}
\end{small}
\begin{small}
and
\begin{verbatim}
TB= 30.0000     M2= 200.000     MU= 200.000     MSQ= 1000.00    
C1=148.714 C2= 266.081 N1= 88.414 N2=152.084 N3= 210.462 N4= 265.541
MST1= 856.771     MST2= 1101.73     MSUL= 975.811     MSUR= 976.706   
MSB1= 973.103     MSB2= 983.796     MSDL= 979.235     MSDR= 977.688   
TAU1= 997.129 TAU2=1004.930 NL= 997.925 EL= 1001.15 ER= 1000.92 
NL1=  997.93

   MHA   CHARGINOS  NEUTRALS   SLEPTONS   SQUARKS  GRAVITINO+GAUGINO
______________________________________________________________________
 1000.00   0.3026    0.1939      0.000      0.000      0.000
\end{verbatim}
\end{small}
Also the pseudoscalar has significant decay rates into charginos and
neutralinos. The charged Higgs branching ratios finally, given in {\tt
br.c1}, {\tt br.c2}, {\tt br.c3} and {\tt br.cs}, are
\begin{small}
\begin{verbatim}
   MHC       BC      TAU NU      MU NU      SU         CS        TB 
______________________________________________________________________
 1002.86 0.7082E-03 0.8290E-01 0.2931E-03 0.8482E-05 0.1583E-03 0.4019
\end{verbatim}
\end{small}
\begin{small}
\begin{verbatim}
   MHC         CD          BU           TS           TD   
______________________________________________________________________
 1002.86   0.6168E-08   0.5140E-05   0.3834E-05   0.1645E-06
\end{verbatim}
\end{small}
\begin{small}
\begin{verbatim}
   MHC         hW           AW        WIDTH 
______________________________________________________________________
 1002.86   0.9836E-05   0.1109E-10    22.77
\end{verbatim}
\end{small}
and
\begin{small}
\begin{verbatim}
TB= 30.0000     M2= 200.000     MU= 200.000     MSQ= 1000.00    
C1=148.714 C2= 266.081 N1= 88.414 N2=152.084 N3= 210.462 N4= 265.541
MST1= 856.771     MST2= 1101.73     MSUL= 975.811     MSUR= 976.706   
MSB1= 973.103     MSB2= 983.796     MSDL= 979.235     MSDR= 977.688   
TAU1= 997.129 TAU2=1004.930 NL= 997.925 EL= 1001.15 ER= 1000.92 
NL1=  997.93

    MHC    CHARG/NEU   SLEPTONS   SQUARKS   GRAVITINO+GAUGINO
______________________________________________________________________
 1002.86    0.5141      0.000      0.000         0.000
\end{verbatim}
\end{small}
The decay into $hW$, although kinematically allowed, is very small,
the one into $AW$ is far off-shell and hence tiny. The decay branching
ratio for the chargino-neutralino final states amounts to more than
50\% and is dominating. Finally the branching ratios of the top quark as
given in the file {\tt br.top} read
\begin{small}
\begin{verbatim}
   MHC         W+- B      H+- B      WIDTH
_______________________________________________________________________________
 1002.86      1.000      0.000      1.336
\end{verbatim}
\end{small}
i.e.~the top quark decays entirely into $Wb$ final states.






\vskip 1 cm
\centerline{\bf Acknowledgments}
\vskip 0.5 cm

The work has been supported in part by the ERC Advanced Grant Higgs@LHC and
the Polish National Science Center HARMONIA 
project under contract UMO-2015/18/M/ST2/00518 (2016-2019). AD would
like to thank CERN for hospitality.

\bibliographystyle{elsarticle-num}
\bibliography{<your-bib-database>}
\hyphenation{Post-Script Sprin-ger}
\providecommand{\href}[2]{#2}\begingroup\raggedright\endgroup






\end{document}